\documentclass[english,onecolumn,fleqn]{svjour2}

\usepackage{amssymb}
\usepackage{amsmath}
\usepackage{mathtools}
\usepackage{epigraph}
\usepackage{eucal}
\usepackage[dvips]{graphics}
\usepackage{epsfig}
\usepackage{bm}
\usepackage{emerald}
\usepackage{babel}
\usepackage{slantsc}
\usepackage{array}
\usepackage{hyperref}
\usepackage[numbers,sort&compress]{natbib}

\def\st{\scriptstyle}

\def\be{\begin{equation}}
\def\ee{\end{equation}}
\def\ba{\begin{eqnarray}}
\def\ea{\end{eqnarray}}

\def\a{\alpha}
\def\b{\beta}
\def\g{\gamma}     \def\G{\Gamma}
\def\d{\delta}

\def\m{\mu}
\def\n{\nu}

\def\s{\sigma}
\def\t{\tau}

\def\la{\label}
\def\pd{\partial}
\def\le{\left}
\def\ri{\right}

\journalname{General Relativity and Gravitation}
\begin{document}
\title{Optical cavity resonator in an expanding universe} 
\author{Sergei M. Kopeikin} 
\institute{Department of Physics \& Astronomy, University of Missouri, 322 Physics Bldg., Columbia, MO 65211, USA;\\
Siberian State Geodetic Academy, 10 Plakhotny St., Novosibirsk 630108, Russia
\email{kopeikins@missouri.edu}}
\date{Received:  15 August, 2014 / Accepted: 1 December 2014   }
\maketitle

\begin{abstract}
We study the cosmological evolution of frequency of a standing electromagnetic wave in a resonant optical cavity placed to the expanding manifold  described by the Robertson-Walker metric. Because of the Einstein principle of equivalence (EEP), one can find a local coordinate system (a local freely falling frame), in which spacetime is locally Minkowskian. However, due to the conformal nature of the Robertson-Walker metric the conventional transformation to the local inertial coordinates introduces ambiguity in the physical interpretation of the local time coordinate, $X^0$. Therefore, contrary to a common-sense expectation, a straightforward implementation of EEP alone does not allow us to unambiguously decide whether atomic clocks based on quantum transitions of atoms, ticks at the same rate as the clocks based on electromagnetic modes of a cavity. To resolve this ambiguity we have to analyse the cavity rigidity and the oscillation of its electromagnetic modes in an expanding universe by employing the full machinery of the Maxwell equations irrespectively of the underlying theory of gravity. We proceed in this way and found out that the size of the cavity and the electromagnetic frequency experience an adiabatic drift in conformal (unphysical) coordinates as the universe expands in accordance with the Hubble law. We set up the oscillation equation for the resonant electromagnetic modes, solve it by the WKB approximation, and reduce the coordinate-dependent quantities to their counterparts measured by a local observer who counts time with atomic clock. The solution shows that there is a perfect mutual cancellation of the adiabatic drift of cavity's frequency by space transformation to local coordinates and the time counted by the clocks based on electromagnetic modes of cavity has the same rate as that of atomic clocks. We conclude that if general relativity is correct and the local expansion of space is isotropic there should be no cosmological drift of frequency of a standing electromagnetic wave oscillating in the cavity resonator as compared to the frequency of atomic clocks. Continuous comparison of the frequency of the optical cavity resonator against that of atomic clock yields a powerful null test of the local isotropy of the Hubble expansion and the Einstein equivalence principle in cosmology.

\end{abstract}
\section{Introduction}

The present paper deals with the analysis of EEP in cosmology by applying general relativity to analyse the clock comparison experiment. We consider atomic and resonance cavity-based clocks, and compare their rates. Experience has shown that a test of a basic principle such as EEP is generally more powerful if one has an example or a framework in which it is violated. There are many examples of such frameworks for testing EEP -- the TH$\varepsilon\m$ framework of Lightman and Lee \citep{Lightman_Lee_1973}, the $\chi$-$g$ framework of Ni \citep{Ni_1983chi-g,Ni_2013}, the Standard Model Extension (SME) of Kostelecky {\it et al} \citep{Kostelesky_2009PhRvD}, and so on. Using such frameworks, one can see exactly how and under what circumstances a non-null result could be obtained, and one can place limits on various EEP-violating parameters. For example, the details of applying the LL framework to clocks, including cavity clocks, appear in Will's textbook \citep{will_1993}. 

Unfortunately, all previously proposed frameworks are limited to the situations where the background spacetime is flat and none was adapted to analyse EEP in case of an expanding cosmological background described by non-stationary FLRW metric. Moreover, there is no general relativistic analysis of the clock-comparison experiments in cosmological setting. Hence, we even do not know whether the clocks ticks with the same rate in general-relativistic cosmology. One may appeal to standard textbooks like \citep{mtw, will_1993} and claim that due to the validity of EEP in general relativity all clocks will tick with the same rate irrespectively of the nature of background metric. This claim is not a scientific proof because EEP only claims that each manifold admits construction of a local Minkowskian frame along time-like worldline, nothing more. EEP does not make a unique prediction on how the local coordinate time is related to physical time counted by clocks. Physical time is generated by a specific physical process which is described by equations of mathematical physics governing the operation of clock's mechanism. These equations should be solved to predict an outcome of a particular clock-comparison experiment. As far as we know, no such a prediction exists in literature regarding the rate of clocks in an expanding universe even in general relativistic framework not mentioning TH$\varepsilon\m$, $\chi$-$g$, and other formalisms. The present paper contributes to the solution of this problem. After it is solved in general relativity, it will be more clear how to describe the clock-comparison experiments in cosmological spacetimes of alternative theories of gravity. 

Evolution of electromagnetic field in an optical cavity resonator with time-dependent geometric parameters was a matter of peer theoretical research in a large number of papers starting from a classic paper by Moore \citep{Moore_1970JMP}. Most of the theoretical papers explored vibrating cavities, where oscillations of cavity walls are periodic in time (see \citep{1996PhRvA..53.2664D,1998PhRvA..57.4784J,Wegrzyn_2007JPhB}, and references therein). Moreover, it was always assumed that the spacetime geometry is flat with no attention paid to the fact that we live in an expanding universe which geometric properties are described by the Friedmann-Leme\^itre-Robertson-Walker (FLRW) metric tensor. Making use of flat geometry looked natural as the Einstein principle of equivalence postulates that tangent spacetime to any curved manifold is locally flat. This postulates is applicable in general relativity but may be invalidated in alternative theories of gravity \citep{2011rcms.book.....K}. In case of an expanding conformally flat manifold the Einstein principle of equivalence (EEP) establishes a connection between the global cosmic time and the local (proper) time measured with the help of clock. There are two types of the most precise clocks - atomic and cavity resonators. They have different principles of operation. Atomic clocks are based on the orbital motion of electrons in atoms while the resonant cavity operates with standing electromagnetic waves. It is pedagogically useful to clarify the connection between the local time measured by the atomic clock and the resonant cavity and the global time of the FLRW universe. 

A common-sense expectation would be that clocks in the expanding universe tick at the same rate, whether they be based on quantum transitions of atoms, or on electromagnetic modes of a cavity. However, (1) there is no textbook on general relativity which addresses this basic point with sufficient mathematical rigour, and (2) direct mathematical transformation from global to local inertial coordinates does not allow us to make a definite conclusion about the rate of clocks based on electromagnetic oscillations in a cavity. We explain the origin and physical nature of this ambiguity in the interpretation of the local time coordinate on expanding FLRW manifold in next section. It motivates us to abandon the counter-intuitive coordinate transformation technique and to switch discussion of the operational principles of resonance-cavity clocks on the basis of more fundamental Maxwell equations in curved spacetime. 

Geometrical coupling between the local physics and the global expansion of the universe has been studied in a fair number of papers (see, for example, reviews \citep{2012PhRvD..86f4004K,2010RvMP...82..169C} and references therein). Most of the previous authors focused on examination of the local gravitational dynamics in the expanding universe while less attention was paid to the cosmological influence on electromagnetic systems like atoms or a local evolution of vacuum electromagnetic fields. Several decades ago this kind of problem looked completely scholastic without any chance to be tested experimentally. Since that time the situation changed dramatically. Atomic clocks, lasers, nanotechnology overturned the world of experimental physics which is operating now fairly close to the fundamental limitations imposed on the measurement of various physical parameters by the quantum field theory \citep{Braginsky_2005PhyU,Braginsky_2007PNAS,Ligo_2011NatPh}. The enormous progress in the experimental techniques stimulates the search for new tests of fundamental gravitational physics both in laboratory and in space. This paper discusses the time evolution of a free electromagnetic wave in a resonant cavity in the conformally-flat metric of the expanding FLRW universe. This problem directly relates to testing the EEP in expanding universe by the clock comparison technique which was pioneered by Turneaure {\it et al} (see \citep{Stein_1978,Turneaure_1983PhRvD}, and references therein).

EEP states that local physics decouples from global phenomena if the effects of spacetime curvature can be neglected. Therefore, it seems natural to expect that if we employ general relativity in the problem under consideration any kind of clock will tick with the same rate. We have checked this expectation in section \ref{lcoeep} by applying EEP to transform the global to local Minkowskian coordinates of a freely falling (Hubble) observer. It turns out that due to the conformal nature of the FLRW metric, the interpretation of the local coordinate time, $X^0$, becomes depending on the worldline along which it is measured. In particular, $X^0/c$ coincides with the proper time $\t$ measured by the observer with an atomic clock but $X^0/c=\t+(1/2)H\t^2$ on the light-like hupersurface of electromagnetic phase, where $H$ is the Hubble constant, and $c$ is the speed of light (see \eqref{ok8c}). Hence, it is not a priori obvious that the clocks based on electromagnetic modes of cavity resonator will have a uniform rate and count the proper time $\t$ with the same frequency as \eqref{ni50} shows. For this reason, even in general relativity the application of EEP alone to particular manifolds like FLRW spacetime, does not warranty the unique answer of clock-comparison experiment, and we have to resort to a more sophisticated mathematical technique to derive the oscillation equation governing the operational procedure of time generation by the cavity clocks in order to compare it with the proper time $\t$ counted by atomic clocks.

Electromagnetic field in a cavity resonator is governed by vacuum (homogeneous) Maxwell's equation. The field freely propagates in cavity in the form of non-interacting plane waves. A linear superposition of these waves form a set of standing waves with frequencies being determined by the cavity length. The cavity is made of a solid, usually an ultra-low expansion glass \citep{2010JaJAP..49f0209M,Tobar5168185} or sapphire \citep{2005RScI...76i4704C}, which strength and stiffness are determined exclusively by the electromagnetic bonding between its atomic and/or molecular components governed by the Coulomb interaction. Therefore, in order to answer the question about how the cavity responds to the cosmological expansion we have to analyse how the Coulomb force is affected by the presence of the time-dependent scale factor, $R=R(t)$, in the FLRW metric. The frequencies form a discrete spectrum of electromagnetic oscillations which is calculated by solving the boundary value problem. It is possible to tune one of the frequencies of the cavity resonance modes to the optical band so that it can be used to interrogate the quantum transitions of atoms placed at the center of the cavity. The electromagnetic interaction between the atoms and the oscillating electromagnetic field allows us to make the most precise clock ever existed \citep{Stein_1978,2004ESASP.554..625G}.

The frequency stability of such a clock is governed by the cavity resonator playing a role of a flywheel oscillator \citep{Tobar5168185}. Long-term stability of the clock is determined by the stability of the frequency of the quantum transitions between the different energy levels of electrons in the atoms used in the clock. The frequency instability of the cavity resonance mode is induced mainly by thermal, acoustic, mechanical, and seismic noise while the fundamental limit is due to the Brownian motion of cavity's reflecting walls \citep{2004PhRvL..93y0602N}. The cavity frequency stability is rapidly improving by making use of innovative technological designs that allows us to reduce the coupling of the cavity to the environmental disturbances. Hence, one can expect that it will be comparable or even exceeds the stability of the atomic transitions \citep{2010OptCo.283.4696Z,2009OExpr..17.8970Z}.

The fractional frequency stability of clocks is characterized by the Allan standard deviation, $\sigma_y(\tau)$, which behaviour crucially depends on the spectrum of the environmental noise \citep{1972pmc..book..466A}. If the spectrum, $S(f)$, of the noise is given in the form of a power law, $S(f)\sim f^\alpha$, the Allan deviation \citep{1972pmc..book..466A} is a power law function as well, $\sigma_y(\tau)\sim \tau^\mu$, where the $\alpha-\mu$ correspondence is well-established \citep{1972pmc..book..466A}. In the ideal case of a white frequency noise the Allan deviation, $\sigma_y(\tau)=\sigma_0/\sqrt{\tau}$, where $\sigma_0$ is the value of the deviation taken at time $\tau=1$ s. Current value of $\sigma_0\sim 10^{-15}$ while the white noise of measuring errors dominates over the time interval 100$\div$1000 s. While the Allan variance is used to measure noise spectra, it also depends on a linear drift of oscillator's frequency. More specifically, if the frequency drift is given by  a linear function of time, $\omega(\tau)=\omega_0+\dot\omega_0\tau$, the Allan deviation $\sigma_y(\tau)=\dot\omega_0\tau/\sqrt{2}$ where the overdot denotes a total  time derivative. The linear frequency drift can mimic the low-frequency noise with the power-law spectrum $S(f)\sim f^{-5}$ \citep{1999MNRAS.305..563K} that is also characteristic for the stochastic gravitational wave background affecting the orbital motion of binary pulsars \citep{kopeikin_1997PhRvD,ilyasov_1998AstL}.

As we show in next section, the Hubble expansion of the universe might cause the frequency $\omega$ of the resonance mode of cavity to drift with $\dot\omega_0/\omega_0= H$, where $ H=2.3\times 10^{-18}$ Hz, is the current value of the Hubble constant estimated from cosmological observations \citep{wmap_2011}. If this drift, induced by the Hubble expansion, is real it reaches after 1000 s the value of $2.3\times 10^{-15}$, and might be measurable. This paper provides a theoretical proof (being independent of the direct application of EEP) that neither frequency of atomic transitions in ions nor the resonance frequency of electromagnetic oscillations in cavity, experience the effect of the Hubble expansion if the global properties of spacetime are described by the FLRW metric. This is in agreement with the EEP stating that any geometric manifold is locally flat. Therefore, our approach gives a new insight to the Einstein equivalence principle on expanding, conformal manifold from the point of view of time metrology. It also opens a new experimental possibilities for testing anisotropy of the Hubble expansion with the help of clocks.

We describe the background geometry of the problem and its relation to local gravitational physics in section \ref{backm}. Application of EEP, local Minkowskain coordinates and difficulties of the interpretation of the Minkowskian time are discussed in section \ref{lcoeep}. Maxwell electrodynamics in FLRW spacetime is outlined in section \ref{meun}. Mechanical rigidity of cavity in an expanding spacetime is established in section \ref{rigres}. The evolution of electromagnetic field and the oscillatory equation of the resonance cavity clocks are derived in section \ref{elemosc}. Concluding remarks are given in section \ref{conrem}. Notations are explained in the course of the presentation.

\section{Background manifold}\la{backm}

We accept that the background manifold is given by the FLRW metric, $ g_{\a\b}$, which describes homogeneous and isotropic spacetime admitting a curved three-dimensional space characterized by a constant parameter $k=\{-1,0,1\}$ \citep{weinberg_1972}, and the Greek indices takes values $0,1,2,3$. Cosmological observations consistently indicate that $k=0$ \citep{wmap_2011,wmap_2013}, and we accept this value in the present paper. The background metric with $k=0$ is 
\be\la{i3}
ds^2=-c^2dt^2+R^2(t)\le(dx^2+dy^2+dx^2\ri) \;, \ee 
where, $c=299 792 458$ m$\cdot$s$^{-1}$, is the fundamental speed, $R(t)$ is the scale factor depending only on the cosmic time $t$, and $x^i=\{x,y,z\}$ are spatial coordinates. Metric (\ref{i3}) is definitely valid on large scales. Local inhomogeneities in the distribution of dark matter may cause the cosmological metric to expand anisotropically \citep{andersson_coley_2011CQG} in different directions so that on small scales the background metric looks like 
\be\la{pm3d}
ds^2=-c^2dt^2+R^2_x(t)dx^2+R^2_y(t)dy^2+R^2_z(t)dz^2\;,
\ee
where $R_x(t), R_y(t), R_z(t)$ are the different scale factors along $x, y, z$ axes respectively. We do not discuss this particular situation in the present paper in full details. Some comments will be given in the last section of the paper. From now on we assume the metric has form (\ref{i3}).

A continuous set of fiducial observers having fixed values of the space coordinates form the Hubble flow for which $dx=dy=dz=0$. Each Hubble observer measures the proper time $\t$ defined by the invariant relation, $ds^2=-c^2d\t^2$. For the fixed spatial coordinates, $\t=t$, which allows us to derive time $t$ from the proper time $\t$ that is measured with the help of an ideal clock. A standard assumption is that the proper time $\t$ can be practically accessed with the help of atomic clocks. This assumption is almost evident in case of an asymptotically flat spacetime. However, in case of an expanding universe this assumption should be proved by deriving equations of motion of an electron in atom and demonstrating that the resulting equations are the same as in asymptotically-flat space time. We provide such a proof in section \ref{rigres}. In what follows, we work in the coordinate system with a Hubble observer placed at its origin, and accept the identity $\t=t$.

For doing further calculations it is convenient to introduce the conformal coordinates $x^\a=(c\eta,x,y,z)$. The conformal time $\eta$ is not directly measurable and relates to the measurable time $\t=t$ by an ordinary differential equation 
\be\la{i4} \frac{d\t}{R(\t)}=d\eta\;. \ee 
The interval in the conformal coordinates is 
\be\la{i5} ds^2=a^2(\eta)\le(-c^2d\eta^2+\d_{ij}dx^idx^j\ri)\;,
\ee
where the scale factor $a(\eta)\equiv R[\t(\eta)]$, the repeated indices in (\ref{i5}) mean the Einstein summation rule, and $\d_{ij}={\rm diag}(1,1,1)$ is the unit matrix. Metric (\ref{i5}) is conformally-equivalent to the Minkowski metric, $\eta_{\a\b}={\rm diag}(-1,+1,+1,+1)$.

FLRW spacetime manifold with the metric (\ref{i5}) has the Christoffel symbols, 
\be\la{i6} \G^\a{}_{\b\g}=-\frac{H}c\le(\d^\a_\b u_\g+\d^\a_\g u_\b- u^\a g_{\b\g}\ri)\;, \ee
where $u^\a=a^{-1}\d^\a_0$ is a four-velocity of the Hubble flow, $ u_\a= g_{\a\b} u^\b$, and $\d^\a_\b$ is the Kroneker symbol (the unit matrix). The Ricci tensor 
\be\la{i6a}  R_{\a\b}=\frac{\dot H}{c^2}\le( g_{\a\b}-2 u_\a u_\b\ri)+3\frac{H^2}{c^2} g_{\a\b}\;, \ee 
and the dot again denotes a derivative with respect to time  $t$.
The dynamic evolution of the FLRW manifold is determined by Einstein's equations 
\be\la{i7}  R_{\a\b}-\frac12 g_{\a\b} R=\frac{8\pi G}{c^4} T_{\a\b}\;, \ee 
where $R=R^\m{}_{\m}=g^{a\b}R_{\a\b}$ is the Ricci scalar and $T_{\a\b}$ is the cumulative energy-momentum tensor of matter (radiation, dust, dark matter and dark energy) governing the dynamic evolution of the universe. The energy-momentum tensor is taken in the form of the perfect fluid and is given by
\be\la{i7bb}  T_{\a\b}=\le(\epsilon+ p\ri) u_\a u_\b+ p g_{\a\b}\;, \ee 
where $\epsilon$ is the energy density, $p$ is pressure, and $u^\a$ is four-velocity of the Hubble flow.
We assume that equations (\ref{i7}) can be solved for the given (\ref{i7}), (\ref{i7bb}), and the time evolution of the scale factor $R(\t)$ is well-defined, at least, in principle. Fortunately, we can get enough valuable physical information on the problem under consideration without knowing the exact solution for $R(\t)$ as we demonstrate in next sections.

\section{Local Minkowskian coordinates in cosmology}\la{lcoeep}

According to the Einstein equivalence principle the cosmological metric (\ref{i3}) should be locally diffeomorphic to the Minkowski metric. Indeed, the diffeomorphism can be achieved by making the following transformation to local Minkowskian coordinates $X^\a=(X^0,X^i)$, where $X^i=(X,Y,Z)$, 
\be\la{i5a}
X^0=c\t+\frac{Hr^2}{2c}+O\le(H^2\ri)\qquad,\qquad X^i=R(\t)x^i+O(H^2)\;,
\ee
where $r^2=\d_{ij}X^i X^j$, $H=\dot R/R$ is the Hubble parameter, the dot over $R$ denotes a derivative with respect to the proper time $\t$, and we have omitted all terms of the order of the time derivative of $H$ and $H^2$ as being negligibly small (notice that according to the Friedmann equations $\dot H\sim H^2$ \citep{weinberg_1972}). There are different types of local coordinates on the FLRW manifold (see, for example, \citep{2005ESASP.576..305K,2006PhRvD..74f4019C,hongya:1920,hongya:1924}) but all of them have the same structure (\ref{i5a}) in the linearised Hubble approximation, when terms of the order of $O(H^2)$ and $O(\dot H)$ are neglected. 

The interval (\ref{i3}) in the local coordinates reads
\be\la{i5b}
ds^2=-\le(dX^0\ri)^2+\delta_{ij}dX^idX^j\;,
\ee 
and all terms of the order of $O(H^2)$ and $O(\dot H)$ have been omitted as playing no role in our discussion due to their exceedingly small observational values. We have now to make connection to physics by establishing a link between the Minkowskian time coordinate, $X^0$, and the time measured by clocks. 
It is tempting to identify the Minkowskian time $X^0$ with the proper time $\t$ of the Hubble observer everywhere in a close neighbourhood of the origin of the local coordinates limited by the condition $r\ll c\sqrt{T_{\rm obs}/H}$ where $T_{\rm obs}$ is the overall time of observation. This is what usually accepted by gravitational physicists \citep{2007CQGra..24.5031M,2010RvMP...82..169C} who use the normalized Minkowskian time, $X^0/c$, as a physical time $\t$ measured with atomic clock. 

We emphasize, however, that such an identification of $X^0/c$ and $\t$ is valid solely on the worldline of the central Hubble observer located at the origin of the local coordinates with $X^i=0$ as immediately follows from (\ref{i5a}). Outside of the worldline of the central observer the Minkowskian time $X^0=X^0(\t,{\bm X})$ is a function of the spatial coordinates ${\bm X}$ which differs from the physical time $\t$ if the clock is not at the origin of the coordinates, and the difference between $X^0/c$ and $\t$ grows with distance proportionally to $Hr^2$. This difference makes impossible a uniform parametrization of worldlines of moving particles in the local coordinates with respect to the physical time $\t$. 

We illustrate it in case of the light rays as the most important physical situation in time metrology of clocks based on propagation of electromagnetic waves in resonance cavity. The light cone of the local Minkowski metric (\ref{i5b}) is given by equation $ds=0$, which has a simple solution describing a light ray moving straight with a uniform speed $c$ with respect to the Minkowski time $X^0$, that is
\be\la{by3c}
X^i=k^iX^0\;,
\ee
where $k^i$ is a unit vector ($|{\bm k}|=\d_{ij} k^i k^j=1$) pointing in the direction of propagation of the light ray. However, on the light cone, the Minkowski time $X^0$ can be calculated from \eqref{i5a}, where we can use an approximation $X^i=ck^i\t$ in the second term being proportional to the Hubble constant $H$. It yields
\be\la{ok8c}
X^0=c\le(\t+\frac12 H\t^2\ri)\;,
\ee
which makes $X^0/c$ apparently different from the proper time $\t$ of the central observer counted along worldline $X^i=0$. This difference leads to 
a non-uniform motion of light rays measured with respect to the proper time $\t$ of the central observer
\be\la{n7g}
X^i=ck^i\le(\t+\frac12 H\t^2\ri)\;.
\ee
which brings about an interesting local effect of the blue shift of the frequency of a freely-propagating radiowave, $\d\omega/\omega=H\t$  \citep{2012PhRvD..86f4004K}. 

Straightforward transformation of Maxwell equations from the global, $x^\a$, to local, $X^\a$, coordinates reduce them to special-relativistic form in accordance with EEP \citep{mtw,will_1993}. However, the conformal nature of the background FLRW manifold does not allow us to identify the Minkowski time $X^0/c$ measured along light rays with the proper time $\t$ measured by atomic clocks. Indeed, a plane-wave solution of Maxwell's equations for the vector potential, $A^\a$, written in the local coordinates, is
\be\la{i8n3}
A^\a_{\pm}={\cal A}^\a\cos\left[\phi_0+\frac1c\left(\omega_0 X^0\pm{\bm K}\cdot{\bm X}\right)\ri]\;,
\ee
where ${\cal A}^\a$ is the constant amplitude, $\omega_0$ is a constant frequency, ${\bm K}=\omega_0{\bm k}$ is the wave vector, ${\bm k}$ is a unit vector along propagation of the wave, $\phi_0$ is the phase constant, and the $\pm$ sign corresponds to two waves traveling in opposite directions. The argument of the cosine function in \eqref{i8n3} is an electromagnetic phase that is a light-like hypersurface made of light rays. Therefore, the time coordinate, $X^0$, entering \eqref{i8n3} is taken on the light ray and given by \eqref{ok8c} in terms of the time $\t$ measured by atomic clocks.

A standing wave in a resonance cavity is a linear superposition of the two plane waves \eqref{i8n3} moving in opposite directions, ${\cal A}^\a={\cal A}^\a_++{\cal A}^\a_-$. The standing wave solution describing the oscillatory behaviour of the electromagnetic modes in cavity is, then,
\be\la{n8c3}
A^\a=2{\cal A}\cos\left[\int_{\t_0}^\t\omega(\t)d\t\ri]\cos\left(\frac{\omega_0}c{\bm k}\cdot{\bm X}\ri)\;,
\ee
where the frequency 
\be\la{ni50}
\omega(\t)=\omega_0(1+H\t)\;,
\ee
is a linearly growing function of time, and the constant phase $\phi_0$ has been absorbed to the lower limit of the integral. 

This is the result that we get by applying EEP on expanding FLRW manifold to propagating electromagnetic waves. Solution \eqref{n8c3}, \eqref{ni50} tells us that we should expect the quadratic-in-time drift of clocks based on electromagnetic modes in cavity with respect to atomic clocks but it does not comply with the expectation engraved in standard textbooks on relativity \citep{mtw,will_1993}. In order to check whether the result \eqref{ni50} is correct we checked several clock-comparison experiments \citep{Stein_1978,Turneaure_1983PhRvD,Storz_1998OptL} which were sensitive enough to detect the linear drift of frequency \eqref{ni50} of electromagnetic modes of a resonance cavity with respect to atomic clock. Those experiments did not detect any difference between the rate of atomic and resonator cavity clocks. The disagreement between theoretical result \eqref{ni50} and the experiments forces us to explore more carefully the operational principle of the resonance cavity-based clock in expanding spacetime manifold, to deeper understand EEP in general relativity and to avoid possible pitfalls in the interpretation of the local Minkowskian coordinates in cosmology.  

\section{Maxwell Electrodynamics in FLRW Universe}\la{meun}

The Maxwell electrodynamics is the most conveniently formulated in terms of the vector potential $A_\a$ of an electromagnetic field. The electromagnetic field tensor is defined in terms of partial derivatives of the vector potential $A_\a$,
\be\la{ft56}
F_{\a\b}\equiv\pd_\a A_{\b}-\pd_\b A_{\a}\;,
\ee 
where $\pd_\a A_{\b}\equiv\pd A_\b/\pd x^\a$. Maxwell's equations for electromagnetic vector $A^\a$ in a curved FLRW spacetime are \citep{mtw}.
\be\la{at1}
A_\a{}^{|\b}{}_{|\b}-A^\b{}_{|\b\a}- R_{\a\b}A^\b=-\frac{4\pi}{c}J_\a\;,
\ee
where the vertical bar and an index after it mean a covariant derivative with respect to the corresponding coordinate, $J_\a$ is a four-vector of a conserved electric current density. It conserves,
\be\la{ggg6}
J^\a{}_{|\a}=0\;,
\ee
as can be derived directly from (\ref{at1}) by taking the covariant divergence from both parts of this equation. 

Misner, Thorne and Wheeler \citep{mtw} recommend to impose on the vector-potential the covariant Lorentz gauge condition, $A^{\b}{}_{|\b}=0$, in order to eliminate the second (gauge-dependent) term in the left side of equation (\ref{at1}) and reduce it to the de Rham wave equation. However, the covariant Lorentz gauge condition does not cancel in (\ref{at1}) the term with the Ricci tensor which makes solution of the de Rham equation problematic. We have found \citep{2012PhRvD..86f4004K} that in case of the FLRW metric both the second and third terms in the left side of (\ref{at1}) can be eliminated if we use a more suitable gauge,
\be\la{at2} 
A^\b{}_{|\b}=-\frac{2}cH A^\b u_\b\;,
\ee
that is equivalent to
\be\la{at2dd} 
\pd_\b A^\b=+\frac{2}cH A^\b u_\b\;.
\ee
This gauge has been also found by M. Ibison \citep{ibison2011} in the form
\be\la{ibis11}
\eta^{\a\b}\pd_\a A_\b=0\;,
\ee
where $A_\b=g_{\a\b}A^\a$. 
Equivalence of \eqref{at2dd} and \eqref{ibis11} can be easily confirmed after substituting $A^\b=g^{\a\b}A_\b=a^{-2}\eta^{\a\b}A_\b$ to the left side of (\ref{at2dd}).

We substitute (\ref{at2}) into covariant equation (\ref{at1}), take the covariant derivatives and use (\ref{i6a}) to express the Ricci tensor in terms of the four velocity and the Hubble parameter. Remarkably, many cancellations take place, and we arrive to an exact Maxwell equations in FLRW spacetime 
\be\la{at3} \Box A_\a=-\frac{4\pi}{c}a^2J_\a\;,
\ee
where $\Box\equiv\eta^{\a\b}\pd^2/\pd x^\a\pd x^\b$ is the ordinary D'Alembert operator in flat spacetime. Thus,
equation (\ref{at3}) looks similar to Maxwell equations in special relativity in the Minkowski space-time except for the presence of the scale factor, $a\equiv a(\eta)$, that appears in the right side of (\ref{at3}). The reason for the simplicity of (\ref{at3}) is the conformal invariance of electromagnetic field that tells us that the Maxwell equations in a conformally flat space must be equivalent to the Maxwell equations in the Minkowski space \citep{wald}. 

Solution of equation (\ref{at3}) is obtained with a standard technique of a retarded potential, 
\be\la{at3a} 
A_\a(\eta,{\bm x})=\frac1c\int_{\st V}\frac{a^2(s)J_a(s,{\bm x}')d^3x'}{|{\bm x}-{\bm x}'|}\;, 
\ee 
where the retarded time $s=\eta-c^{-1}|{\bm x}-{\bm x}'|$, and the integral is performed over the spatial volume $V$ occupied by the matter charge distribution. Equation (\ref{at3a}) tells us that weak electromagnetic waves propagate with speed $c$ in the conformal coordinates $(\eta,{\bm x})$ of the FLRW universe with the space curvature $k=0$.

The gauge condition (\ref{at2}) applied to equation (\ref{at3a}) is consistent with the fundamental law of conservation of electric current (\ref{ggg6})
which is equivalent to
\be\la{at5}
\frac1c\frac{\pd}{\pd \eta}\le(a^2 J_{0}\ri)-\frac{\pd}{\pd x^i}\le(a^2 J_{i}\ri)=0\;.
\ee
Hence, the total electric charge of the matter distribution
\be\la{at6}
 Q\equiv -\int_{\st V}a^2(\eta)J_0(\eta,{\bm x})d^3x\;,
\ee
is constant, and does not change due to the course of the Hubble expansion, at least, up to terms of the order of $H^2$. 

\section{Rigidity of resonant cavity}\la{rigres}

Before discussing evolution of electromagnetic field in a resonant cavity, we have to make sure that the Hubble expansion does not change the geometric length of the cavity. Rigidity of the cavity is determined by the chemical bonds between atoms. They are governed primarily by the electric Coulomb force acting between atomic nuclei. Hence, we have to explore if the Coulomb law preserves its classic form in the expanding FLRW universe.

The Coulomb force is governed by an electric potential $\phi\equiv cA_0$ of an atomic nucleus. It can be found by making Taylor expansion
of the retarded arguments of the vector potential (\ref{at3a}) around time $\eta$. As the charge is conserved, the expansion term of the order of $1/c$ vanishes, and we arrive to 
\be\la{at3b} \phi(\eta,{\bm x})=\int_{\st V}\frac{a^2(\eta)J_0(\eta,{\bm x}')d^3x'}{|{\bm x}-{\bm x}'|}+O\le(\frac1{c^2}\ri)\;.
\ee
Outside of the charge distribution, $|{\bm x}|>|{\bm x}'|$, and the integral can be expanded in terms of the electric multipoles -- the constant charge $Q$, the dipole electric moment $Q_i$, and so on \citep{Jackson_1998},
\be\la{at7} 
\phi(\eta,{\bm x})=-\frac{Q}{|{\bm x}|}+\frac{Q_i x^i}{|{\bm x}|^3}+...\;.
\ee
For the sake of simplicity, we neglect the dipole and higher-order electric multipoles. Their treatment makes calculations more tedious but do not change the conclusion of the present section on the rigidity of the cavity. 

The orbital motion of an electron in a curved spacetime is given in the main approximation by the second Newton's law with an electromagnetic Lorentz force standing in its right side while the left side of this law is a covariant derivative from the linear momentum of electron $p^\a={m}dx^\a/d\s$, 
\be\la{at8} 
m\le(\frac{d^2x^a}{d\s^2}+\G^\a{}_{\b\g}\frac{dx^\b}{d\s}\frac{dx^\g}{d\s}\ri)=eF^\a{}_{\b}\frac{dx^\b}{d\s}\;,
\ee
where $\s$ is the affine parameter along the electron's orbit, $e$ is the charge of electron $(e<0)$, and ${m}$ is electron's mass. The charge and mass of electron remain constant in expanding universe (see equation (\ref{at6}) and our paper \citep{2012PhRvD..86f4004K} for the proof). The Christoffel symbols in (\ref{at8}) account for the presence of the gravitational field of the FLRW universe. 
If the conformal time $\eta$ is used for parametrization of the worldline of electron, equation (\ref{at8}) assumes the following form 
\be\la{at9} \frac{d^2 x^i}{d\eta^2}+\G^i{}_{\m\n}\frac{dx^\m}{d\eta}\frac{dx^\n}{d\eta}-\G^0{}_{\m\n}\frac{dx^\m}{d\eta}\frac{dx^\n}{d\eta}\frac{dx^i}{d\eta}=\frac{e}{{m}}\le(F^{i}{}_\b-F^0{}_\b\frac{dx^i}{d\eta}\ri)\frac{dx^\b}{d\eta}\frac{d\s}{d\eta}\;.
\ee
In the slow-motion approximation $d\s/d\eta=a(\eta)$, $F^i{}_0=F_{i0}/a^2$, and the electric field of the atomic nucleus is $E_i=F_{i\b}(dx^\b/d\eta)=\pd_i\phi$. By neglecting relativistic corrections of the order of $1/c^2$ in (\ref{at9}), we obtain the equation of motion of electron in the expanding universe 
\be\la{at10} 
\frac{d^2 \bm x}{d\eta^2}=-H\frac{d\bm x}{d\eta}+\frac{eQ}{{m}a}\frac{{\bm x}}{|{\bm x}|^3}\;. 
\ee 
By doing the conformal spacetime transformation 
\be\la{eqo7} \t=\int a(\eta)d\eta\;,\qquad X^i=a(\eta)x^i\;, 
\ee 
we can recast (\ref{at10}) to the classic form of the second Newton's law with Coulomb's force 
\be\la{at11} {m}\frac{d^2{\bm X}}{d\t^2}=\frac{eQ}{\rho^3}{\bm X}\;,
\ee
where $\rho=|{\bm X}|$, the dot means the time derivative with respect to the physical time $\t$ of the central observer.
Thus, the orbital motion of electrons, after having been expressed in physical coordinates $\le(\t,{\bm X}\ri)$ does not reveal any dependence on the scale factor $a(\eta)$ and the Hubble parameter $H$. This proves that the strength of the chemical bonds of the material from which the cavity is made of, is not affected by the expansion of universe. Hence, one concludes that the cavity is rigid and keeps its physical length constant in the expanding universe under condition that temperature and other environmental disturbances are kept under control and their influence can be subtracted with as high precision as necessary.

\section{Electromagnetic oscillations in a resonant cavity}\la{elemosc}

Electromagnetic field in a resonant cavity obeys the homogeneous Maxwell equations that follows immediately from (\ref{at3}) 
\be\la{eo1} 
\Box A_\a=0\;,
\ee
and we assumed a perfect vacuum inside the cavity (no dielectric). 
The field propagates between the cavity walls in the form of a transverse-traceless electromagnetic wave. We assume that the walls are ideal mirrors reflecting electromagnetic waves without dissipation so that the wave is fully confined in the cavity. We also make a simplification that the cavity is a rectangular box with the spatial coordinate axes $x^i=(x,y,z)$ directed along its sides. It allows us to solve (\ref{eo1}) by applying the method of separation of variables in the Cartesian coordinates that splits equation (\ref{eo1}) in three, one-dimensional wave equations which are decoupled \citep{Jackson_1998}. 

The condition (\ref{at2}) bears a residual gauge freedom for the electromagnetic potential $A_\a\longrightarrow A'_\a=A_\a+\pd_\a\chi$ where $\chi$ is an arbitrary scalar function that obeys a homogeneous wave equation $\Box\chi=0$. 
The residual gauge freedom allows us to chose, $A_0=0$, in the solution of (\ref{eo1}) representing an electromagnetic wave. We further assume that the wave is propagating along $x$-axis and is polarized in the $z$-direction so that the potential $A_i=(0,0,A)$, while the electric $E_\a=F_{\a\b} u^\b$, and the magnetic $B_\a=-(1/2)\epsilon_{\a\b\m\n}F^{\m\n} u^\b$, fields have the following spatial components, 
\be\la{eo4} E_i=(0,0,E)=\le(0,0,-\frac1{a}\frac{\pd A}{\pd\eta}\ri)\;,\qquad B_i=(0,B,0)=\le(0,-\frac1{a}\frac{\pd A}{\pd x},0\ri)\;.
\ee
The field $E\equiv E(\eta,x)$ and $B\equiv B(\eta,x)$ satisfy the wave equation that is derived from (\ref{eo1}). In particular, the wave equation for the electric field, $E=E_z=F_{z0}u^0\equiv{\cal E}/a$, is 
\be\la{eqo5}
-\frac{\pd^2 \cal E}{\pd\eta^2}+\frac{\pd^2\cal E}{\pd x^2}=0\;, 
\ee 
and a similar equation exists for the magnetic field ${\cal B}\equiv B a$. Equation (\ref{eqo5}) could be, of course, derived directly without making use of the vector potential $A_\a$. This approach requires an independent derivation of the covariant equation for $F_{\a\b}$ in the expanding universe that will lead to a complicated de Rham operator for the tensor field of a second rank \citep{koppetr}. Simplification of this operator and its reduction to the wave equation (\ref{eqo5}) is based on a rather long and tedious chain of mathematical transformations which will be published somewhere else.

The tangential components of $E_i$ must vanish on the cavity's walls that are perpendicular to $x$-axis. The cavity keeps its geometric shape unchanged in the local physical coordinates $X^i=a(t)x^i$ but it is expanding adiabatically in accordance with the Hubble law in the conformal coordinates $x^i$. Let as chose the left wall of the cavity as the origin of $x$ and $X$ axis. The right wall of the cavity is fixed in the local coordinates at $X=L$. Then, the right wall of the cavity is moving adiabatically with respect to the conformal coordinates $x^i$ as $x=l(\eta)=L/a(\eta)$. 

The boundary conditions imposed on the electric field at the cavity's walls are ${\cal E}[\eta,x=0]=0$ and ${\cal E}[\eta,x=l(t)]=0$ \citep{Jackson_1998}.
Solution of (\ref{eqo5}) with the boundary condition depending on time has been worked out in a number of papers   \citep{Moore_1970JMP,1994PhRvL..73.1931L,1996PhRvA..53.2664D,1998PhRvA..57.4784J}. It consists of a discrete spectrum of standing waves where each harmonic can be represented as a Fourier series with respect to the {\it instantaneous} basis
\be\la{eq6}
{\cal E}(\eta,x)=\sum_k Q_k(\eta)\sin\le[\frac{\pi k x}{l(\eta)}\ri]\;.
\ee 
This form of the solution satisfies the boundary conditions exactly. 

We put (\ref{eq6}) to the wave equation (\ref{eqo5}), multiply it with $\sin\le[\pi n x/l(t)\ri]$ ($n\not= k)$, and integrate over $x$ from $x=0$ to $x=l(t)$. It yields the ordinary differential equation for the amplitude $Q_k$ that reads  
\be\la{eqo12} {Q}''_k+\Omega_k^2(\eta){Q}_k=+2H\sum_{n=1}^\infty w_{kn}\Omega_n{Q}'_n\;, \ee 
where the prime denotes a time derivative with respect to the conformal time $\eta$, the (time-dependent) resonant frequencies 
\be\la{eq12a} \Omega_k(\eta)\equiv a(\eta)\omega_k\;, \ee 
and 
\be\la{eqo13} w_{mn}\equiv \frac2{L}\int^L_0X\sin\le(k_m X\ri)\cos\le(k_n X\ri)dX\;, \ee 
is a set of constant coefficients, $\omega_k\equiv (c\pi k)/L$ is a constant frequency, and $k_m\equiv \pi m/L$,  $k_n\equiv \pi n/L$.

The scale factor $a(\eta)$ changes very slowly due to the Hubble expansion. For this reason, equation (\ref{eq12a}) can be solved in an adiabatic approximation which neglects all terms with the time derivatives of $Q_k$ in the right side of (\ref{eq12a}). The adiabatic solution of (\ref{eqo12}) is
\be\la{eq14}
\sqrt{a(\eta)}Q_k(\eta)=a_k\sin\le[\int\Omega_k(\eta)d\eta\ri]+b_k\cos\le[\int\Omega_k(\eta)d\eta\ri]\;,
\ee
where $a_n$ and $b_n$ are constant amplitudes of the $n$-th mode of the harmonic oscillations.
We notice that in the adiabatic approximation, the product $\int\Omega_k(\eta)d\eta=\omega_k \t$. Therefore, the electric field in the cavity is
\be\la{bv3x1}
E(t,{\bm X})=\sum_k\biggl[A_k\sin\omega_k\t+B_k\cos\omega_k\t\biggr]\sin\le(\frac{\pi k X}{L}\ri)\;,
\ee
where we have used transformation from the global to local spatial coordinates $x=Xa(\eta)$ and introduced notations $A_n\equiv a_n R^{-3/2}(\t)$, $B_n\equiv b_nR^{-3/2}(\t)$ for the adiabatically-changing amplitudes of the orthogonal modes of the oscillations.

Equation (\ref{bv3x1}) demonstrates that the frequency of electromagnetic oscillations in a resonant cavity is not subject to the Hubble expansion because the observed frequencies, $\omega_n$, are constant with respect to the proper time $\t$ of the Hubble observer measured with the help of atomic clock. This theoretical conclusion has a rather precise experimental verification \citep{Storz_1998OptL}. The amplitudes $A_n(\t)$, $B_n(\t)$ of the standing wave in (\ref{bv3x1}) depend on physical time $\t$ due to the Hubble expansion through the time-dependent scale factor $R(\t)$. This change in the amplitude of the electric field is minuscule on any (practically reasonable) observational time span and can be safely neglected. 

\section{Conclusion}\la{conrem}

We have studied evolution of electromagnetic frequency of a standing electromagnetic wave in a rectangular resonant cavity used as a clock in the FLRW universe. We have analysed how EEP works in case of the expanding space and the subtitles in the physical interpretation of the local Minkowskian time. We have  proved that in general relativity the physical time counted by the cavity-based clock has the same rate as the proper time measured by atomic clocks despite of the expanding, non-stationary background metric.  

It is possible to extend our calculations to a cavity of any geometric shape by directly solving Maxwell's equations for oscillating modes in the same way as in section \ref{elemosc}. In any case the frequency of the proper modes of the cavity is not subject to the Hubble expansion in accordance with the Einstein principle of equivalence \citep{2011rcms.book.....K}. This result may not hold in some alternative metric-based theories of gravity which admit existence of other long-range gravitational fields (scalar, vector) besides the metric tensor \citep{2006LRR.....9....3W,2010AIPC.1256....3T}. It can be also violated in the non-metric theories of gravity which admit torsion and non-metricity in the affine connection which violate the Einstein principle of equivalence \citep{Hehl_1976RvMP,Baekler_2011PhRvD,Vitagliano_2014CQGra}. 

We would like to emphasize that the rectangular cavity resonator can experience a relative drift of frequencies of electromagnetic waves oscillating in $x,y,z$ directions respectively if a local anisotropy of the cosmological metric taken in the form of equation \ref{pm3d} is present in nature. Let us denote $H\equiv H_x=\dot R_x/R_x$ the value of the Hubble constant in the direction of $x$-axis, and $H_y=\dot R_y/R_y$ and $H_z=\dot R_z/R_z$ the values of the Hubble constant in the direction of $y$ and $z$ axes respectively.  Then, the differences between the anisotropic scale factors are $\d R_y(\t)\equiv R_y(\t)-R_x(\t)=\d H_y \t$ and $\d R_z(\t)\equiv R_z(\t)-R_x(\t)=\d H_z \t$,  where $\d H_y=H_y-H$ and $\d H_z\equiv H_z-H$. This would lead to a relative drift of frequencies $\omega_y$ and $\omega_z$ of the standing waves in $y$ and $z$ direction with respect to the frequency $\omega\equiv \omega_x$ of oscillation of the wave in $x$ direction,
\be\la{by7cm}
\omega_y=\omega+\d H_y\t\;,\qquad  \omega_z=\omega+\d H_z\t\;.
\ee
This new type of metrological measurement of time by clocks based on the optical cavity resonators belongs to the class of the Hughes-Drever experiments \citep{matt_2005LRR} which test a spatial anisotropy of mass and Mach's principle in special relativity. The experiment described in the present paper goes beyond special relativity and probes one of the most fundamental cosmological assumptions - the local isotropy of the Hubble expansion and the Einstein equivalence principle. 

It should be noticed that here is a long history of comparing different kinds of clocks to look for cosmological variations. The discussion of such experiments is usually done in terms of variations of the fine-structure constant, the gyromagnetic ratio of the proton or the electron/proton mass ratio but not in terms of test of EEP and the possible anisotropy in expansion of space. Microwave resonator cavities have been used in comparison with atomic clocks to place bounds on the Lorentz-violation parameters of SME, but they have not been used to look for cosmological drifts \citep{Wolf_2003PhRvL,Lipa_2003PhRvL,Mueller_2003PhRvL}. Nonetheless, the data from these experiments may be retroactively used for testing EEP in cosmology. 
Further theoretical and experimental exploration of this subject is desirable.

\begin{acknowledgements}
I am grateful to two anonymous referees for valuable remarks and comments that helped to improve the manuscript. This work was supported by the Faculty Fellowship 2014 in the College of Arts and Science of the University of Missouri and the grant 14-27-00068 of the Russian Scientific Foundation.
\end{acknowledgements}
\bibliographystyle{unsrt}
\bibliography{bibliography}

\begin{thebibliography}{10}

\bibitem{Lightman_Lee_1973}
A.~P. {Lightman} and D.~L. {Lee}.
\newblock {Restricted Proof that the Weak Equivalence Principle Implies the
  Einstein Equivalence Principle}.
\newblock {\em \prd}, 8:364--376, July 1973.

\bibitem{Ni_1983chi-g}
W.-T. {Ni}.
\newblock {Equivalence Principles, Their Empirical Foundations, and the Role of
  Precision Experiments to Test Them}.
\newblock In W.-T. {Ni}, editor, {\em Proceedings of the 1983 International
  School and Symposium on Precision Measurement and Gravity Experiment, Taipei,
  Republic of China, January 24-February 2, 1983}, pages 491--517. National
  Tsing Hua University, Hsinchu, Taiwan, Republic of China, 1983.

\bibitem{Ni_2013}
W.-T. Ni, H.-H. Mei, and S.-J. Wu.
\newblock Foundations of classical electrodynamics, equivalence principle and
  cosmic interactions: A short exposition and an upadate.
\newblock {\em Modern Physics Letters A}, 28(03):1340013 [15 pages], 2013.

\bibitem{Kostelesky_2009PhRvD}
V.~A. {Kosteleck{\'y}} and M.~{Mewes}.
\newblock {Electrodynamics with Lorentz-violating operators of arbitrary
  dimension}.
\newblock {\em \prd}, 80(1):015020 [58 pages], July 2009.

\bibitem{will_1993}
C.~M. {Will}.
\newblock {\em {Theory and Experiment in Gravitational Physics}}.
\newblock Cambridge: Cambridge University Press, 396 pp., March 1993.

\bibitem{mtw}
C.~W. {Misner}, K.~S. {Thorne}, and J.~A. {Wheeler}.
\newblock {\em {Gravitation}}.
\newblock San Francisco: W.H.~Freeman and Co., 1973.

\bibitem{Moore_1970JMP}
G.~T. {Moore}.
\newblock {Quantum Theory of the Electromagnetic Field in a Variable-Length
  One-Dimensional Cavity}.
\newblock {\em Journal of Mathematical Physics}, 11:2679--2691, September 1970.

\bibitem{1996PhRvA..53.2664D}
V.~V. {Dodonov} and A.~B. {Klimov}.
\newblock {Generation and detection of photons in a cavity with a resonantly
  oscillating boundary}.
\newblock {\em \pra}, 53:2664--2682, April 1996.

\bibitem{1998PhRvA..57.4784J}
M.~{Janowicz}.
\newblock {Evolution of wave fields and atom-field interactions in a cavity
  with one oscillating mirror}.
\newblock {\em \pra}, 57:4784--4790, June 1998.

\bibitem{Wegrzyn_2007JPhB}
P.~{Wegrzyn}.
\newblock {Exact closed-form analytical solutions for vibrating cavities}.
\newblock {\em Journal of Physics B Atomic Molecular Physics}, 40:2621--2640,
  July 2007.

\bibitem{2011rcms.book.....K}
S.~{Kopeikin}, M.~{Efroimsky}, and G.~{Kaplan}.
\newblock {\em {Relativistic Celestial Mechanics of the Solar System}}.
\newblock Weinheim: Wiley-VCH, September 2011.

\bibitem{2012PhRvD..86f4004K}
S.~M. {Kopeikin}.
\newblock {Celestial ephemerides in an expanding universe}.
\newblock {\em \prd}, 86(6):064004, September 2012.

\bibitem{2010RvMP...82..169C}
M.~{Carrera} and D.~{Giulini}.
\newblock {Influence of global cosmological expansion on local dynamics and
  kinematics}.
\newblock {\em Reviews of Modern Physics}, 82:169--208, January 2010.

\bibitem{Braginsky_2005PhyU}
V.~B. {Braginsky}.
\newblock {ANNUS MIRABILIS. PHYSICS OF OUR DAYS: Development of quantum
  measurement methods (Methodological notes on part of Einstein's scientific
  legacy)}.
\newblock {\em Physics Uspekhi}, 48:595--600, June 2005.

\bibitem{Braginsky_2007PNAS}
V.~B. {Braginsky}.
\newblock {Inaugural Article: Experiments with probe masses}.
\newblock {\em Proceedings of the National Academy of Science}, 104:3677--3680,
  March 2007.

\bibitem{Ligo_2011NatPh}
{Ligo Scientific Collaboration}, J.~{Abadie}, B.~P. {Abbott}, R.~{Abbott},
  T.~D. {Abbott}, M.~{Abernathy}, C.~{Adams}, R.~{Adhikari}, C.~{Affeldt},
  B.~{Allen}, and et~al.
\newblock {A gravitational wave observatory operating beyond the quantum
  shot-noise limit}.
\newblock {\em Nature Physics}, 7:962--965, December 2011.

\bibitem{Stein_1978}
S.~R. Stein and J.~P. Turneaure.
\newblock Superconducting resonators: High stability oscillators and
  applications to fundamental physics and metrology.
\newblock {\em AIP Conference Proceedings}, 44(1), 1978.

\bibitem{Turneaure_1983PhRvD}
J.~P. {Turneaure}, C.~M. {Will}, B.~F. {Farrell}, E.~M. {Mattison}, and
  R.~F.~C. {Vessot}.
\newblock {Test of the principle of equivalence by a null gravitational
  red-shift experiment}.
\newblock {\em \prd}, 27:1705--1714, April 1983.

\bibitem{2010JaJAP..49f0209M}
M.~{ Koide} and T.~{Ido}.
\newblock {Design of Monolithic Rectangular Cavity of 30-cm Length}.
\newblock {\em Japanese Journal of Applied Physics}, 49(6):060209, June 2010.

\bibitem{Tobar5168185}
J.~Millo, Y.~Le~Coq, S.~Bize, J.~Guena, H.~Jiang, M.~Abgrall, E.M.L. English,
  A.~Clairon, G.~Santarelli, and M.E. Tobar.
\newblock Flywheel oscillator for atomic fountain clocks using ultra-stable
  lasers and a fiber-based optical frequency comb.
\newblock In {\em Frequency Control Symposium, 2009 Joint with the 22nd
  European Frequency and Time forum. IEEE International}, pages 280 --281,
  april 2009.

\bibitem{2005RScI...76i4704C}
D.~{Chambon}, S.~{Bize}, M.~{Lours}, F.~{Narbonneau}, H.~{Marion},
  A.~{Clairon}, G.~{Santarelli}, A.~{Luiten}, and M.~{Tobar}.
\newblock {Design and realization of a flywheel oscillator for advanced time
  and frequency metrology}.
\newblock {\em Review of Scientific Instruments}, 76(9):094704, September 2005.

\bibitem{2004ESASP.554..625G}
P.~{Gill}, G.~P. {Barwood}, H.~A. {Klein}, G.~{Huang}, H.~S. {Margolis}, S.~N.
  {Lea}, M.~{Oxborrow}, and S.~A. {Webster}.
\newblock {Optical clocks and ultra-stable optical oscillators for navigation,
  space science and astronomy}.
\newblock In B.~{Warmbein}, editor, {\em 5th International Conference on Space
  Optics}, volume 554 of {\em ESA Special Publication}, pages 625--630, June
  2004.

\bibitem{2004PhRvL..93y0602N}
K.~{Numata}, A.~{Kemery}, and J.~{Camp}.
\newblock {Thermal-Noise Limit in the Frequency Stabilization of Lasers with
  Rigid Cavities}.
\newblock {\em Physical Review Letters}, 93(25):250602, December 2004.

\bibitem{2010OptCo.283.4696Z}
Y.~N. {Zhao}, J.~{Zhang}, J.~{Stuhler}, G.~{Schuricht}, F.~{Lison}, Z.~H. {Lu},
  and L.~J. {Wang}.
\newblock {Sub-Hertz frequency stabilization of a commercial diode laser}.
\newblock {\em Optics Communications}, 283:4696--4700, December 2010.

\bibitem{2009OExpr..17.8970Z}
Y.~N. {Zhao}, J.~{Zhang}, A.~{Stejskal}, T.~{Liu}, V.~{Elman}, Z.~H. {Lu}, and
  L.~J. {Wang}.
\newblock {A vibration-insensitive optical cavity and absolute determination of
  its ultrahigh stability}.
\newblock {\em Optics Express}, 17:8970, May 2009.

\bibitem{1972pmc..book..466A}
D.~W. {Allan}.
\newblock {Statistics of atomic frequency standards}.
\newblock In {\em Precision Measurement and Calibration. Selected NBS Papers on
  Frequency and Time.}, volume~5 of {\em NBS Special Publication 300}, page
  466. Washington, DC: US Goverment Printing Office, June 1972.

\bibitem{1999MNRAS.305..563K}
S.~M. {Kopeikin}.
\newblock {Millisecond and binary pulsars as nature's frequency standards - II.
  The effects of low-frequency timing noise on residuals and measured
  parameters}.
\newblock {\em \mnras}, 305:563--590, May 1999.

\bibitem{kopeikin_1997PhRvD}
S.~M. {Kopeikin}.
\newblock {Binary pulsars as detectors of ultralow-frequency gravitational
  waves}.
\newblock {\em \prd}, 56:4455--4469, October 1997.

\bibitem{ilyasov_1998AstL}
Y.~P. {Ilyasov}, S.~M. {Kopeikin}, and A.~E. {Rodin}.
\newblock {The astronomical timescale based on the orbital motion of a pulsar
  in a binary system}.
\newblock {\em Astronomy Letters}, 24:228--236, March 1998.

\bibitem{wmap_2011}
N.~Jarosik, C.~L. Bennett, J.~Dunkley, B.~Gold, M.~R. Greason, M.~Halpern,
  R.~S. Hill, G.~Hinshaw, A.~Kogut, E.~Komatsu, D.~Larson, M.~Limon, S.~S.
  Meyer, M.~R. Nolta, N.~Odegard, L.~Page, K.~M. Smith, D.~N. Spergel, G.~S.
  Tucker, J.~L. Weiland, E.~Wollack, and E.~L. Wright.
\newblock Seven-year wilkinson microwave anisotropy probe (wmap) observations:
  Sky maps, systematic errors, and basic results.
\newblock {\em The Astrophysical Journal Supplement Series}, 192(2):14, 2011.

\bibitem{weinberg_1972}
S.~{Weinberg}.
\newblock {\em {Gravitation and Cosmology: Principles and Applications of the
  General Theory of Relativity}}.
\newblock New York: John Wiley \& Sons, Inc., July 1972.

\bibitem{wmap_2013}
C.~L. Bennett, D.~Larson, J.~L. Weiland, N.~Jarosik, G.~Hinshaw, N.~Odegard,
  K.~M. Smith, R.~S. Hill, B.~Gold, M.~Halpern, E.~Komatsu, M.~R. Nolta,
  L.~Page, D.~N. Spergel, E.~Wollack, J.~Dunkley, A.~Kogut, M.~Limon, S.~S.
  Meyer, G.~S. Tucker, and E.~L. Wright.
\newblock Nine-year wilkinson microwave anisotropy probe (wmap) observations:
  Final maps and results.
\newblock {\em The Astrophysical Journal Supplement Series}, 208(2):20, 2013.

\bibitem{andersson_coley_2011CQG}
L.~{Andersson} and A.~{Coley}.
\newblock {EDITORIAL: Inhomogeneous cosmological models and averaging in
  cosmology: overview Inhomogeneous cosmological models and averaging in
  cosmology: overview}.
\newblock {\em Classical and Quantum Gravity}, 28(16):160301, August 2011.

\bibitem{2005ESASP.576..305K}
S.~A. {Klioner} and M.~H. {Soffel}.
\newblock {Refining the Relativistic Model for Gaia: Cosmological Effects in
  the BCRS}.
\newblock In C.~{Turon}, K.~S. {O'Flaherty}, and M.~A.~C. {Perryman}, editors,
  {\em The Three-Dimensional Universe with Gaia}, volume 576 of {\em ESA
  Special Publication}, pages 305--308, January 2005.

\bibitem{2006PhRvD..74f4019C}
C.~{Chicone} and B.~{Mashhoon}.
\newblock {Explicit Fermi coordinates and tidal dynamics in de Sitter and
  G{\"o}del spacetimes}.
\newblock {\em \prd}, 74(6):064019, September 2006.

\bibitem{hongya:1920}
Liu Hongya.
\newblock {Cosmological models in globally geodesic coordinates. I. Metric}.
\newblock {\em Journal of Mathematical Physics}, 28(8):1920--1923, 1987.

\bibitem{hongya:1924}
Liu Hongya.
\newblock {Cosmological models in globally geodesic coordinates. II. Near-field
  approximation.}
\newblock {\em Journal of Mathematical Physics}, 28(8):1924--1927, 1987.

\bibitem{2007CQGra..24.5031M}
B.~{Mashhoon}, N.~{Mobed}, and D.~{Singh}.
\newblock {Tidal dynamics in cosmological spacetimes}.
\newblock {\em Classical and Quantum Gravity}, 24:5031--5046, October 2007.

\bibitem{Storz_1998OptL}
R.~{Storz}, C.~{Braxmaier}, K.~{J{\"a}ck}, O.~{Pradl}, and S.~{Schiller}.
\newblock {Ultrahigh long-termdimensional stability of a sapphire cryogenic
  optical resonator}.
\newblock {\em Optics Letters}, 23:1031--1033, July 1998.

\bibitem{ibison2011}
M.~{Ibison}.
\newblock {The Dirac Field at the Future Conformal Singularity}.
\newblock In A.~{Ghribi}, editor, {\em Advances in Modern Cosmology}, pages 139
  -- 172. InTech, 2011.

\bibitem{wald}
R.~M. {Wald}.
\newblock {\em {General relativity}}.
\newblock Chicago: University of Chicago Press, 1984.

\bibitem{Jackson_1998}
J.~D. {Jackson}.
\newblock {\em {Classical Electrodynamics, 3rd Edition}}.
\newblock John Wiley \& Sons, Inc.: New York, USA, July 1998.

\bibitem{koppetr}
S.~M. {Kopeikin} and A.~N. {Petrov}.
\newblock {Post-Newtonian celestial dynamics in cosmology: Field equations}.
\newblock {\em \prd}, 87(4):044029, February 2013.

\bibitem{1994PhRvL..73.1931L}
C.~K. {Law}.
\newblock {Resonance response of the quantum vacuum to an oscillating
  boundary}.
\newblock {\em Physical Review Letters}, 73:1931--1934, October 1994.

\bibitem{2006LRR.....9....3W}
C.~M. {Will}.
\newblock {The Confrontation between General Relativity and Experiment}.
\newblock {\em Living Reviews in Relativity}, 9:3, March 2006.

\bibitem{2010AIPC.1256....3T}
S.~G. {Turyshev}.
\newblock {Testing General Relativity in the Solar System: Present Status and
  Possible Future Developments}.
\newblock In H.~A. {Morales-Tecotl}, L.~A. {Urena-Lopez}, R.~{Linares-Romero},
  and H.~H. {Garcia-Compean}, editors, {\em American Institute of Physics
  Conference Series}, volume 1256 of {\em American Institute of Physics
  Conference Series}, pages 3--26, July 2010.

\bibitem{Hehl_1976RvMP}
F.~W. {Hehl}, P.~{von der Heyde}, G.~D. {Kerlick}, and J.~M. {Nester}.
\newblock {General relativity with spin and torsion: Foundations and
  prospects}.
\newblock {\em Reviews of Modern Physics}, 48:393--416, July 1976.

\bibitem{Baekler_2011PhRvD}
P.~{Baekler}, F.~W. {Hehl}, and J.~M. {Nester}.
\newblock {Poincar{\'e} gauge theory of gravity: Friedman cosmology with even
  and odd parity modes: Analytic part}.
\newblock {\em \prd}, 83(2):024001, January 2011.

\bibitem{Vitagliano_2014CQGra}
V.~{Vitagliano}.
\newblock {The role of nonmetricity in metric-affine theories of gravity}.
\newblock {\em Classical and Quantum Gravity}, 31(4):045006, February 2014.

\bibitem{matt_2005LRR}
D.~{Mattingly}.
\newblock {Modern Tests of Lorentz Invariance}.
\newblock {\em Living Reviews in Relativity}, 8:5, September 2005.

\bibitem{Wolf_2003PhRvL}
P.~{Wolf}, S.~{Bize}, A.~{Clairon}, A.~N. {Luiten}, G.~{Santarelli}, and M.~E.
  {Tobar}.
\newblock {Tests of Lorentz Invariance using a Microwave Resonator}.
\newblock {\em Physical Review Letters}, 90(6):060402, February 2003.

\bibitem{Lipa_2003PhRvL}
J.~A. {Lipa}, J.~A. {Nissen}, S.~{Wang}, D.~A. {Stricker}, and D.~{Avaloff}.
\newblock {New Limit on Signals of Lorentz Violation in Electrodynamics}.
\newblock {\em Physical Review Letters}, 90(6):060403, February 2003.

\bibitem{Mueller_2003PhRvL}
H.~{M{\"u}ller}, S.~{Herrmann}, C.~{Braxmaier}, S.~{Schiller}, and A.~{Peters}.
\newblock {Modern Michelson-Morley Experiment using Cryogenic Optical
  Resonators}.
\newblock {\em Physical Review Letters}, 91(2):020401, July 2003.

\end{thebibliography}
\end{document}